\documentclass[aps,prl,twocolumn,footinbib,superscriptaddress,longbibliography,amsmath,amssymb]{revtex4-2} 

\usepackage{graphicx}
\usepackage{subfigure}
\usepackage{epsfig} 
\usepackage{dcolumn}
\usepackage{bm}
\usepackage{times} 
\usepackage{amsmath}
\usepackage{float}
\usepackage{color}
\usepackage{multirow}
\usepackage[normalem]{ulem}

\begin{document}
\title{Specular-Andreev reflection and Andreev interference in an Ising superconductor junction}

\author{Gaoyang Li}
\affiliation{Graduate School of China Academy of Engineering Physics, Beijing 100193, China}
\author{Sourabh Patil}
\affiliation{Fachbereich Physik, Universit\"{a}t Konstanz, D-78457 Konstanz, Germany}
\author{Yanxia Xing}
\affiliation{Key Laboratory of Advanced Optoelectronic Quantum Architecture and Measurement, Ministry of Education, Beijing Institute of Technology, Beijing 100081, China}
\author{Wolfgang Belzig}
\affiliation{Fachbereich Physik, Universit\"{a}t Konstanz, D-78457 Konstanz, Germany}
\author{Gaomin Tang}
\email{gmtang@gscaep.ac.cn}
\affiliation{Graduate School of China Academy of Engineering Physics, Beijing 100193, China}

\begin{abstract}
Being resilient to magnetic field, Ising superconductor serves as an exceptional platform for studying the interplay between superconductivity and magnetism. In this Letter, we first explore the transport properties of a two-terminal graphene–Ising superconductor junction where mirage gaps are induced in the superconductor by an exchange field due to magnetic proximity effect. We demonstrate that the chemical potential range of graphene supporting specular-Andreev reflection at the interface is between the two mirage gaps and about twice the Ising spin-orbit coupling strength. This enhances the resilience of observing specular-Andreev reflection against graphene potential fluctuations in experiments. We further study the Andreev interference effect based on a four-terminal junction of which two terminals consist of Ising superconductors in the presence of exchange fields. 
Due to the finite contribution from the spin-triplet pairing, the interference can be modulated by tuning the relative orientation of the exchange fields in addition to the traditional scheme by changing superconducting phase difference and the chemical potential of the normal region.
\end{abstract}

\maketitle

{\it Introduction.}
Ising superconductivity, recently observed in monolayer transition-metal dichalcogenides such as MoS$_2$ and NbSe$_2$, presents a unique platform to explore the interplay between superconductivity and magnetism~\cite{Ye12, Taniguchi12, Ye15, Saito16, Xi16}. Due to the absence of inversion symmetry in these materials, valley-dependent spin-orbit coupling emerges, pinning electron spins out-of-plane with opposite orientations at the $K$ and $K'$ valleys. This intrinsic Ising spin-orbit coupling significantly enhances the resilience of superconductivity against in-plane magnetic fields. Furthermore, applying an in-plane magnetic field induces two mirage gaps, positioned symmetrically away from the main superconducting gap at energies approximately equal to the Ising spin-orbit coupling strength~\cite{mirage, ising_patil, ilic_mirage_2023}. In addition to conventional spin-singlet pairing, Ising superconductors also support equal-spin triplet pairing states, suggesting unique pairing mechanisms~\cite{Mockli19, Mockli20, Wickramaratne20}. Moreover, charge and spin transport phenomena in van der Waals junctions based on Ising superconductors have recently attracted significant research interest, both theoretically~\cite{Zhou16, Transport_Sun18, Transport_Sun19, GT21, Transport_Sun22_1, Transport_Sun22_2, Asano24, He24, GT24} and experimentally~\cite{Idzuchi21, Jeon21, Kang22, Zalic23, Xiong24}.

At the interface between a metal and a superconductor, electrons incident from the metallic side under a DC voltage bias can undergo Andreev reflection, where they are reflected as holes. In addition to the conventional retro-Andreev reflection (RAR), specular-Andreev reflection (SAR) in which the incident electrons and reflected holes reside in the conduction and valence bands, respectively, was first proposed in the context of graphene-based superconducting junctions~\cite{Specular_06}. 
The signature of SAR is identified as a distinct conductance dip, which occurs when the applied bias voltage aligns with the chemical potential of graphene~\cite{Specular_06, Ram23}.
However, observing SAR in monolayer graphene-superconductor junctions is challenging due to strong potential fluctuations near the Dirac point~\cite{Efetov2016}.
This obstacle has been overcome by using bilayer graphene, where the enhanced density of states near the Dirac point significantly suppresses potential fluctuations~\cite{Efetov2016, Efetov16, Ludwig07, Takane17}.  

In this Letter, we first investigate the RAR and SAR in a two-terminal graphene–Ising superconductor junction. By introducing an exchange field from an adjacent ferromagnetic layer, mirage gaps are induced in the Ising superconductor. We demonstrate that both the main superconducting gap and the mirage gaps support SAR, facilitating its experimental observation. Additionally, we explore an Andreev interferometer implemented in a four-terminal junction, where two superconducting electrodes modulate electron interference within a graphene nanoribbon. We show that both the RAR and SAR, occurring at the main and mirage gaps, contribute to the formation of Andreev interference patterns. These patterns exhibit a sensitive dependence on the superconducting phase difference and the relative orientation of the exchange fields in the superconducting electrodes. The dependence of the patterns on these parameters is analytically derived using the scattering matrix method and subsequently confirmed by numerical calculations based on the nonequilibrium Green's function formalism.

{\it Ising superconductivity.}
The Bogoliubov-de Gennes Hamiltonian of an Ising superconductor with an s-wave pairing gap $\Delta$
near the $K$ or $K'$ valley in the Nambu basis $(c_{\bm{p},\uparrow}, c_{\bm{p},\downarrow}, c_{-\bm{p},\uparrow}^\dag, c_{-\bm{p},\downarrow}^\dag)^T$
reads~\cite{Zhou16, GT21}
\begin{equation}
  H_{\rm BdG}(\bm{p},s) = 
  \begin{bmatrix}
    H_0({\bm{p}, s})  &  \Delta i\sigma_y  \\
    -\Delta i\sigma_y  &  -H_0^*(-\bm{p}, -s) 
  \end{bmatrix}.
\end{equation}
The Hamiltonian $H_0$ is 
\begin{equation}
  H_0(\bm{p},s)=\xi_{\bm{p}}\sigma_0 + s\beta_{\rm so} \sigma_z - \bm{J}
  \cdot \bm{\sigma} ,
\end{equation}
where $\bm{p}$ is the momentum deviation from $\bm{K}$ or $\bm{K}'$, $s=\pm 1$ denotes the valley index, and $\xi_{\bm{p}}$ is the dispersion measured from the chemical potential of the superconductor. The Pauli matrices $\sigma_x, \sigma_y$, and $\sigma_z$ act on the spin space, with $\sigma_0$ the corresponding unit matrix. The Ising spin-orbit coupling strength is denoted as $\beta_{\rm so}$. The Zeeman term $\bm{J} \cdot \bm{\sigma}$ arises from the in-plane exchange field $\bm{J}$ provided by the adjacent ferromagnetic layer. The eigenvalues $E$ of the Hamiltonian are obtained from
\begin{equation}
  E^2 = \xi_{\bm{p}}^2 + J_{\rm eff}^2 + \Delta^2 \pm 2 \sqrt{ \xi_{\bm{p}}^2 J_{\rm eff}^2 + J^2 \Delta^2},
\end{equation}
with $J_{\rm eff}^2=J^2+\beta_{\rm so}^2$. Due to the presence of the exchange field, the main gap is reduced to $2\Delta_{\rm eff}$ with $\Delta_{\text{eff}} = \beta_{so} \Delta / J_{\rm eff}$. Additionally, two symmetric mirage gaps emerge at energies $\varepsilon_0=\pm (\varepsilon_1+\varepsilon_2)/2$, with $\varepsilon_{1(2)}=\sqrt{J_{\rm eff}^2+\Delta^2\pm 2J\Delta}$~\cite{mirage,ising_patil}.

\begin{figure}
\includegraphics[width=\columnwidth]{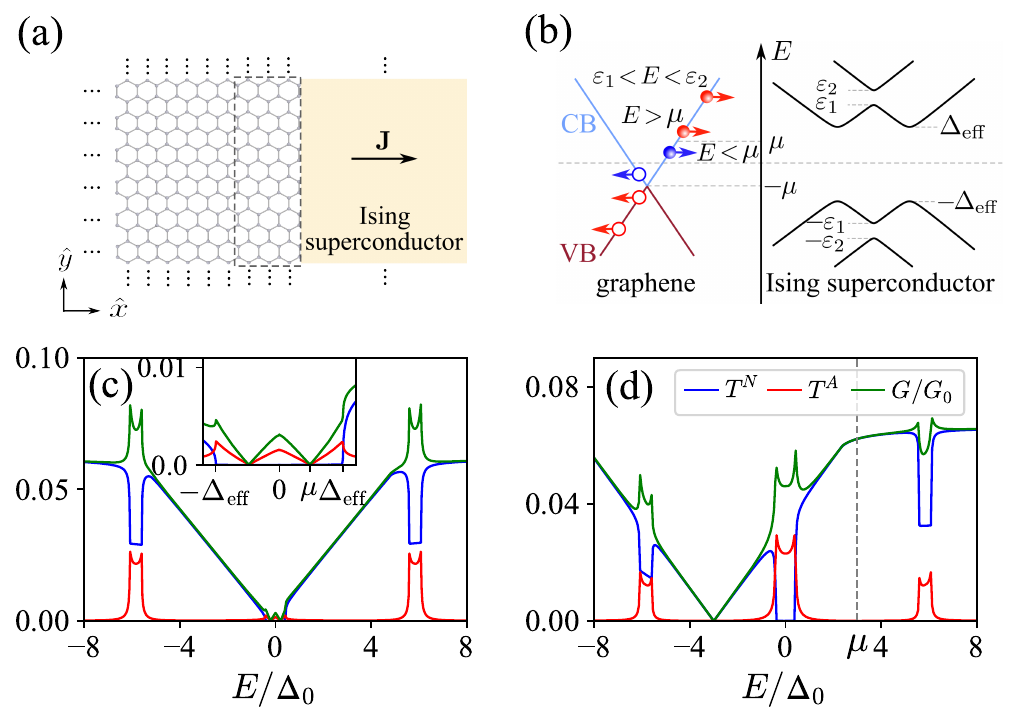}
\caption{(a) Schematic of a two-terminal graphene-Ising superconductor junction with translational symmetry along the $y$ direction. (b) Illustration of Andreev reflections at the interface. The mirage gap edges are denoted by $\pm\varepsilon_1$ and $\pm\varepsilon_2$ in addition to the main gap of $2\Delta_{\rm eff}$. The chemical potential of graphene $\mu$ lies within the main gap. Incident electrons from the conduction band (CB) with energies $0<E<\mu$ (blue solid circle) undergo retro-reflection as holes in the same band (blue open circle). Electrons with energies $\mu < E < \Delta_{\rm eff}$ or $\varepsilon_1<E<\varepsilon_2$ (red solid circles) are specularly reflected in the valence band (VB) (red open circles). Normal transmission $T^N$, Andreev reflection coefficient $T^A$ and differential conductance $G/G_0$ at (c) $\mu = 0.2\Delta_0$ and (d) $\mu = 3\Delta_0$, where $G_0$ is the conductance quantum. The effective main gap is $\Delta_{\rm eff} = 0.42\Delta_0$ and the mirage-gap edges are $\pm\varepsilon_1=\pm 5.6\Delta_0$ and $\pm\varepsilon_2=\pm 6.1\Delta_0$. }
\label{fig1}
\end{figure}

{\it Specular-Andreev reflection.}
We begin by investigating the Andreev reflections in a two-terminal system consisting of a zigzag graphene nanoribbon coupled to an Ising superconductor under an in-plane exchange field $\bm{J}$ [see Fig.~\ref{fig1}(a)]. 
The exchange field gives rise to mirage gaps, in which both RAR and SAR can take place. Specifically, within these gaps, including the main and mirage gaps, RAR occurs for $|E|<|\mu|$ and SAR for $|E|>|\mu|$. 
Here, $E$ represents the energy of the incident electron, and $\mu$ is the chemical potential of graphene.
Consequently, the chemical potential range for observing the SAR is  $0<|\mu|<\varepsilon_2$, which is broader than the range reported in Ref.~\cite{Specular_06}.
Figure~\ref{fig1}(b) illustrates the case where $\mu$ lies within the main gap.
Due to strong potential fluctuations in graphene relative to the superconducting gap, experimentally maintaining the chemical potential within the range required to observe SAR is challenging~\cite{Efetov2016}. However, the emergence of mirage gaps broadens the range of chemical potential favorable for SAR, reducing the experimental difficulty in observing this phenomenon.

We further employ numerical calculations to investigate the Andreev reflections and differential conductances.
The tight-binding Hamiltonian of graphene in the Nambu basis $\psi_i=(c_{i\uparrow}, c_{i\downarrow}, c_{i\uparrow}^\dag, c_{i\downarrow}^\dag)^T$ is expressed as
\begin{equation}\label{TB1}
  H = -t\sum_{\langle ij\rangle} \psi_i^\dag \tau_3 \otimes \sigma_0 \psi_j -\mu \sum_i \psi_i^\dag \tau_3\otimes \sigma_0 \psi_i,
\end{equation}
where $i$ and $j$ indicate the lattice sites, $t=2.75\,$eV is the nearest-neighbor hopping energy, $\mu$ is the chemical potential, the third Pauli matrix $\tau_3$ acts on the Nambu space, and $\otimes$ denotes the Kronecker product. Due to translational symmetry along the $y$ direction, the Hamiltonian in Eq.~\eqref{TB1} can be decomposed as $H = \sum_{k_y} H(k_y)$ where $H(k_y)$ is the one-dimensional Hamiltonian for a given $k_y$. Using the nonequilibrium Green's function formalism, the normal transmission and Andreev reflection coefficients are, respectively, calculated by~\cite{Specular_Sun_09, Sun24}
\begin{align}
  T^N &= \sum_{k_y,s} {\rm tr}\big[ \Gamma_{Lee} \big(G^r \Gamma_{R} G^a)_{ee}\big],\label{TN} \\
  T^A &= \sum_{k_y,s} {\rm tr}( \Gamma_{Lee}G_{eh}^r\Gamma_{Rhh}G_{he}^a),\label{TA}
\end{align}
where the trace is taken over the spin and site indices. The subscripts $L$ and $R$, respectively, label the graphene and superconducting electrodes, and $e(h)$ denotes the electron (hole) degree of freedom. The linewidth function is given by $\Gamma_\alpha = -2{\rm Im} \left[\Sigma_\alpha^r(k_y)\right]$ with $\alpha = L, R$, where $\Sigma_\alpha^r(k_y)$ is the retarded self-energy due to the coupling between the central region and electrode $\alpha$. The self-energy of the graphene electrode is calculated numerically using the transfer-matrix method~\cite{TransferMatrix1,TransferMatrix2}, and that of the superconducting electrode is given in Ref.~[\onlinecite{SM}]. The retarded Green's function is obtained as $G^r(k_y) = \big[E - H(k_y) - \sum_\alpha \Sigma_\alpha^r(k_y)\big]^{-1}$. Finally, the differential conductance is $G(V) = [T^N(eV) + 2T^A(eV)] G_0$ with $G_0 = 2e^2/h$ the conductance quantum.

In the numerical calculation, we set the critical temperature of the superconductor to be $T_c = 8\,$K. The zero-temperature superconducting gap is $\Delta_0 = 1.76 k_B T_c = 1.22\,$meV in the absence of external fields. The Ising spin-orbit coupling strength is fixed at $\beta_{\rm so} = 5 \Delta_0$, the exchange field is $J = 3 \Delta_0$, and the temperature is $T = 0.01 T_c$. The Dynes broadening parameter is chosen to be $\eta = 0.01 \Delta_0$. The order parameter $\Delta$ is determined self-consistently~\cite{mirage}.

Figures~\ref{fig1}(c) and \ref{fig1}(d) show the normal transmission, Andreev reflection, and differential conductance for different graphene chemical potentials $\mu$. When $\mu$ lies within the main gap, SAR occurs for energies $E$ within both the main and the mirage gaps [see Fig.~\ref{fig1}(c)].
Notably, we observe enhanced SAR in the mirage gaps compared to the Andreev reflections in the main gap. This enhancement arises because the density of states in graphene increases linearly with energy.
For the same reason, when $\mu$ lies within the continuous spectrum region with $\Delta_{\rm eff} <|\mu|<\varepsilon_1$, the Andreev reflections in Fig.~\ref{fig1}(d) are enhanced in the main gap but reduced in the mirage gaps compared to those in Fig.~\ref{fig1}(c). Additionally, the normal transmission coefficients in the mirage gaps are finite due to the finite density of states in Ising superconductors.

\begin{figure}
\includegraphics[width=\columnwidth]{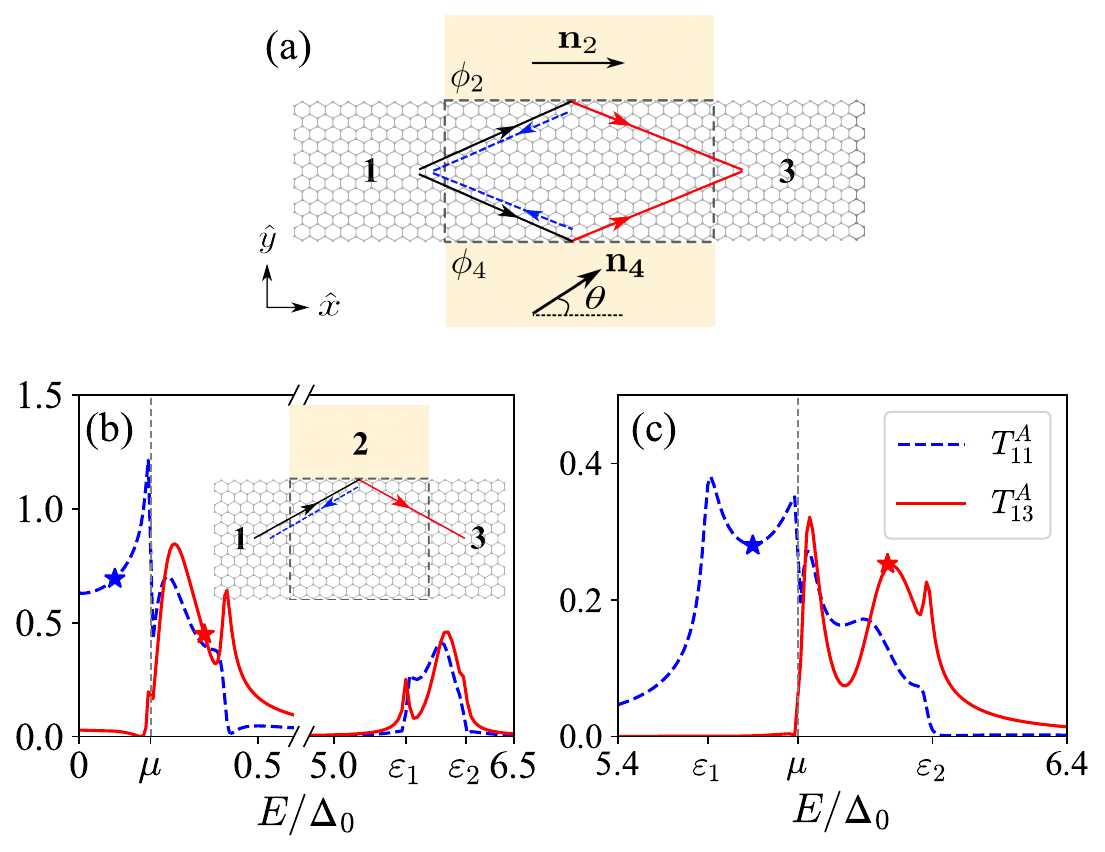}
\caption{(a) Schematic of an Andreev interferometer comprising a central region (boxed area) connected to two Ising superconductors (electrodes $2$ and $4$) and two metallic electrodes (electrodes $1$ and $3$). The central region facilitates both retro-Andreev (dashed blue arrows) and specular-Andreev (solid red arrows) reflections. 
The interference patterns can be tuned by the relative angle $\theta$ between the exchange fields in the superconductors and the superconducting phase difference $\phi=\phi_4-\phi_2$. The central region and metallic electrodes are modeled using a zigzag graphene nanoribbon. The schematic shows a central graphene region of size $6\times 17$, while the numerical calculations are performed for a region of $80\times 80$. 
The retro-Andreev and specular-Andreev reflections for the three-terminal configuration [see inset of (b)] are shown at (b) $\mu=0.2\Delta_0$ and (c) $\mu=5.8\Delta_0$. Other parameters are the same as those in Figs.~\ref{fig1}(c) and (d).
}
\label{fig2}
\end{figure}

{\it Andreev interference.}
We further design an Andreev interferometer based on a four-terminal junction with two Ising superconductor electrodes and two metallic electrodes. 
As illustrated in Fig.~\ref{fig2}(a), the central region and electrodes $1$ and $3$ are composed of zigzag graphene nanoribbon, while electrodes $2$ and $4$ are Ising superconductors with a superconducting phase difference $\phi=\phi_4-\phi_2$. 
The Andreev-reflected holes from electrodes $2$ and $4$ interfere at electrodes $1$ and $3$. 
Unlike Ref.~\cite{Specular_Sun_09}, our setup introduces an additional control parameter, namely, the relative angle $\theta$ between the exchange fields in the superconducting electrodes. Furthermore, Andreev reflections occurring within the mirage gaps can also be harnessed to construct and modulate the interference patterns.

We first analytically derive the Andreev interference coefficients at electrodes $1$ and $3$ using the scattering matrix formalism. 
In this approach, we adopt the infinite-interface approximation, in which the graphene-superconductor interfaces are assumed infinitely wide. This simplification neglects boundary-induced diffraction effects arising from finite-size contacts in the four-terminal geometry. The RAR and SAR are characterized by processes in which electrons injected from electrode $1$ are reflected as holes to electrodes $1$ and $3$, respectively. The Andreev-reflected holes originating from the two superconducting electrodes interfere coherently at electrodes $1$ and $3$, generating measurable interference patterns.
The resulting total electron-hole scattering matrices are given by
\begin{equation} \label{r11_13}
  \bm{r}_{11(13)}^A = \bm{r}_{eh, 2} \pm \bm{r}_{eh, 4},
\end{equation}
where $\bm{r}_{eh, \alpha}$ is the electron-hole scattering matrix associated with superconducting electrode $\alpha$. 
The minus sign in Eq.~\eqref{r11_13} for $\bm{r}_{13}^A$ arises from the odd parity of SARs~\cite{Specular_Sun_11, Parity_odd_09, Beenakker08}. 
The interference coefficients can be obtained from $T_\beta^A = 2 {\rm tr} \big(\bm{r}_{\beta}^A {\bm{r}_{\beta}^A}^\dag \big)$ with $\beta=11, 13$ denoting the interference at electrodes $1$ and $3$, respectively. The factor of $2$ accounts for the valley degrees of freedom. 

For incident energies within the main gap, the electron-hole scattering matrix for electrode $\alpha$ is given by $\bm{r}_{eh, \alpha} = \exp(i\phi_\alpha + i\chi + i \gamma \bm{n}_\alpha \cdot \bm{\sigma})$ with $\alpha=2,4$ for superconducting electrodes, and $\chi = \arccos(E /\Delta_{\text{eff}})$~\cite{GT24}. Here, $\gamma$ is the precession angle with $\sin\gamma = J / J_{\rm eff}$. The total interference coefficients in the main gap are explicitly expressed as
\begin{align}
  T_{11}^A &= 4(1+\cos\phi) + 4(1 -\cos\phi + 2 \cos\theta \cos\phi) \sin^2\gamma , \label{T11A}\\
  T_{13}^A &= 4(1-\cos\phi) + 4(1 +\cos\phi - 2 \cos\theta \cos\phi) \sin^2\gamma \label{T13A}.
\end{align}
In the absence of exchange fields, the interference coefficients simplify to $T_{11(13)}^A=4(1\pm \cos\phi)$, demonstrating a phase-tunable interference. 
In the presence of exchange fields, we have $T_{11}^A|_{\phi = \pi}=T_{13}^A|_{\phi = 0} = 4(1- \cos\theta)J^2/J_{\rm eff}^2$. These coefficients reach their maximum value of $8J^2/J_{\rm eff}^2$ at $\theta=\pi$ and vanishes at $\theta=0$. For $\phi = \pm \pi/2$, both $T_{11}^A$ and $T_{13}^A$ are independent of $\theta$, taking the value $T_{11}^A = T_{13}^A = 2+2J^2/J_{\rm eff}^2$.

For incident energies within the mirage gaps, spin-triplet pairing correlations dominate~\cite{mirage,Silaev}. By taking $\xi_{\bm{p}}=0$, the pairing-correlation function can be approximated as
\begin{equation}
  F_\alpha(s,E) \approx \Delta e^{i\phi_\alpha}\big[ F_x(s,E)\sigma_x + F_y(s,E)\sigma_y\big] i\sigma_y,
\end{equation}
where $F_x$ and $F_y$ are provided in the Supplemental Material~\cite{SM}. 
Since the matrix $\bm{r}_{eh,\alpha}$ is proportional to the pairing-correlation function $F_\alpha(s,E)$, the interference coefficients at electrode $1$ and electrode $3$ are, respectively, given by
\begin{align} 
  & T_{11}^A \propto \Delta^2 J^2(E^2+\beta_{\rm so}^2 )(1+\cos\phi\cos\theta) / M^2 ,  \label{T11Am}
  \\
  & T_{13}^A \propto \Delta^2 J^2(E^2+\beta_{\rm so}^2 )(1-\cos\phi\cos\theta) / M^2 , \label{T13Am}
\end{align}
with
$M = (E^2-\Delta^2+J_{\rm eff}^2)^2 -4E^2 J_{\rm eff}^2 +4\beta_{\rm so}^2 \Delta^2$.

\begin{figure}
\includegraphics[width=\columnwidth]{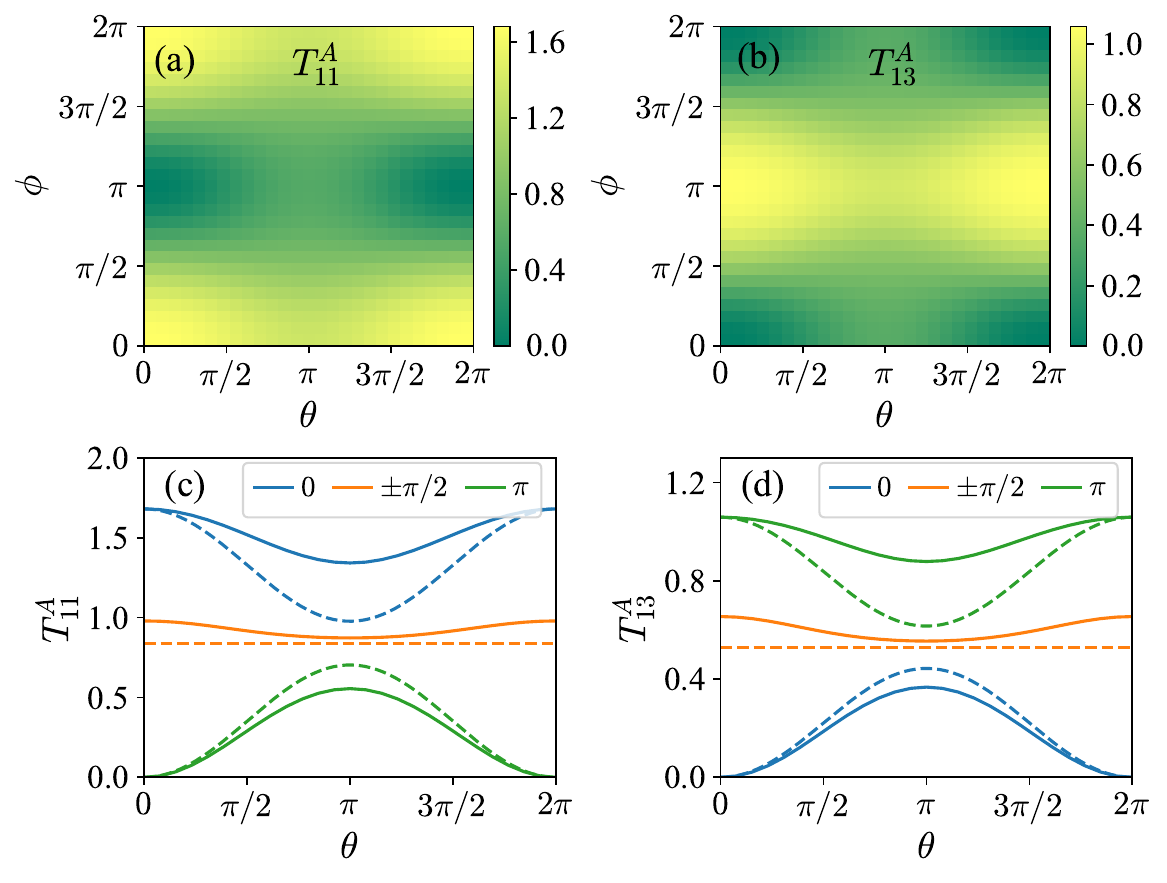}
\caption{Interference patterns within the main gap with chemical potential $\mu=0.2\Delta_0$. (a) Interference coefficient $T_{11}^A$ at energy $E=0.1\Delta_0$ [blue star in Fig.~\ref{fig2}(b)] and (b) interference coefficient $T_{13}^A$ at $E=0.35\Delta_0$ [red star in Fig.~\ref{fig2}(b)]. Solid lines in (c) and (d) are the line-cuts of the interference patterns for various superconducting phase differences in (a) and (b), respectively.
The dashed lines in (c) and (d), scaled by factors of $0.17$ and $0.11$ respectively, correspond to the analytical results given by Eqs.~\eqref{T11A} and \eqref{T13A}.
}
\label{fig3}
\end{figure}

To validate the analytical predictions, we numerically calculate the interference coefficients $T_{11}^A$ and $T_{13}^A$ using the lattice Green's function method. Before exploring the interference effects, we first analyze the Andreev reflections in a three-terminal configuration, as shown in the inset of Fig.~\ref{fig2}(b), which differs from 
Fig.~\ref{fig2}(a) by the absence of electrode $4$. Its symmetric counterpart (without electrode $2$) exhibits identical Andreev reflection amplitudes, except for an additional $\pi$ phase shift in the SARs due to their odd parity~\cite{Specular_Sun_11, Parity_odd_09, Beenakker08}.
The Andreev reflection coefficient from electrode $1$ to electrode $\alpha$ is given by
\begin{equation}
  T_{\rm 1\alpha}^A =\sum_s {\rm tr}( \Gamma_{1ee}G_{eh}^r\Gamma_{\alpha hh}G_{he}^a), 
\end{equation}
where $\alpha = 1,3$ denotes the graphene electrodes.
Figures~\ref{fig2}(b) and \ref{fig2}(c) illustrate $T_{11}^A$ and $T_{13}^A$ versus incident energy $E$ for graphene chemical potential $\mu$ within the main and mirage gap, respectively. Pronounced peaks are observed at the edges of the main gap and the mirage gaps.
Note that in the two-terminal system with translational symmetry as shown in Fig.~\ref{fig1}(a), RAR occurs for $|E|<|\mu|$, while SAR occurs for $|E|>|\mu|$. 
In contrast, in the three-terminal system, Andreev-reflected holes undergo significant diffraction due to the finite-size effects, resulting in a mixture of $T_{11}^A$ and $T_{13}^A$. 
With increasing the size of the central region, $T_{11}^A$ at $|E| > |\mu|$ and $T_{13}^A$ at $|E| < |\mu|$ decreases~\cite{Specular_Sun_09}.

\begin{figure}
\includegraphics[width=\columnwidth]{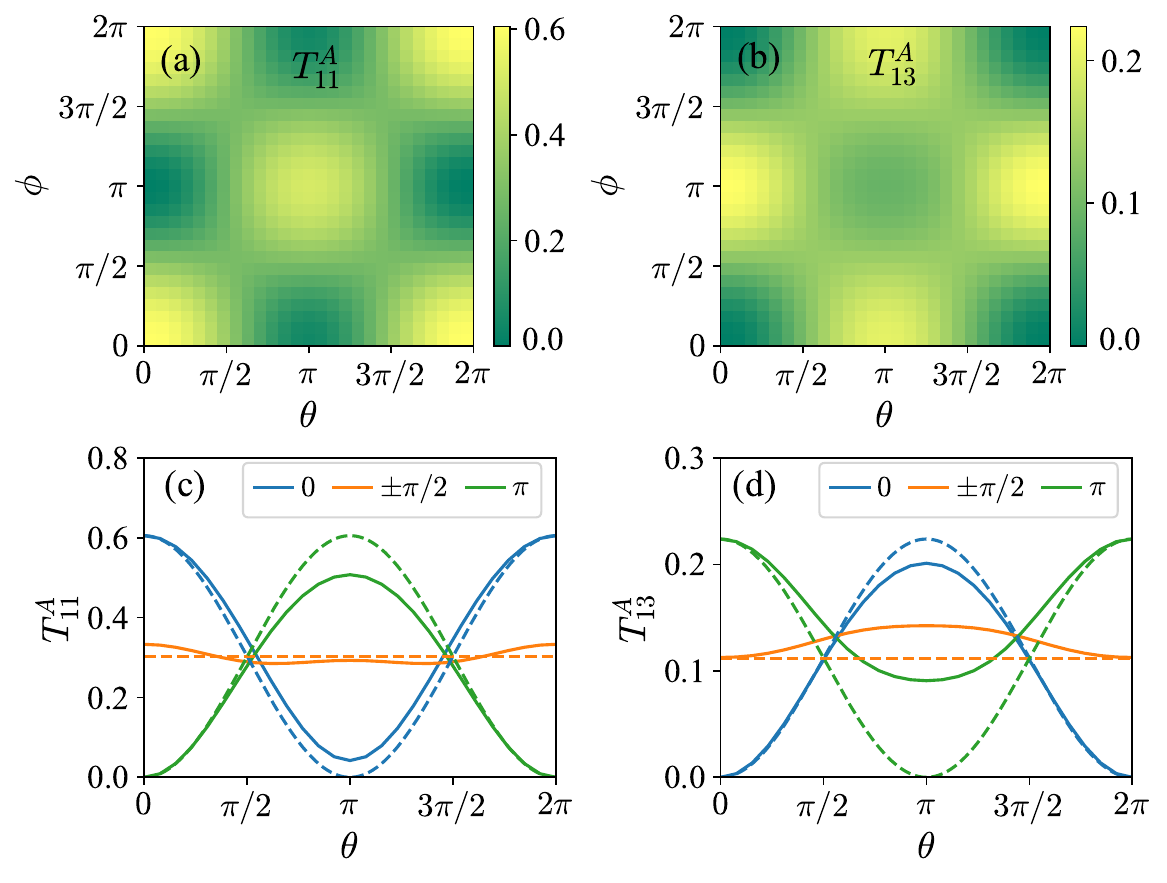}
\caption{Interference patterns within the mirage gaps with chemical potential $\mu=5.8\Delta_0$. (a) Interference coefficient $T_{11}^A$ at energy $E=5.7\Delta_0$ [blue star in Fig.~\ref{fig2}(c)] and (b) interference coefficient $T_{13}^A$ at energy $E=6\Delta_0$ [red star in Fig.~\ref{fig2}(c)]. Solid lines in (c) and (d) are the line-cuts of the interference patterns for various superconducting phase differences in (a) and (b), respectively. The dashed lines in (c) and (d), scaled by factors of $0.3$ and $0.11$, respectively, correspond to $1\pm \cos\phi\cos\theta$ from Eqs.~\eqref{T11Am} and \eqref{T13Am}.
}
\label{fig4}
\end{figure}

In Fig.~\ref{fig3}, we present the numerical interference patterns for $T_{11}^A$ and $T_{13}^A$ in the four-terminal interferometer, with the chemical potential set within the main gap. 
The incident electron energies in Figs.~\ref{fig3}(a) and \ref{fig3}(b) are below and above the chemical potential $\mu$, respectively. 
For parallel exchange fields, the interference coefficient $T_{11}^A$ reaches its maximum at $\phi = 0$ and is completely suppressed at $\phi = \pi$, while $T_{13}^A$ reaches its maximum at $\phi = \pi$ and is completely suppressed at $\phi = 0$.
Notably, the relative angle of the exchange fields has a sizable impact on the interference when $\phi = 0$ or $\pi$.
This behavior arises from the coexistence of the spin-singlet and spin-triplet parings in the main gap. While adjusting $\theta$ primarily modulates the contribution from the spin-triplet pairings to the interference, it leaves the contribution from the spin-singlet pairing largely unaffected. Consequently, within the main gap, the interference shows a relatively weaker dependence on the relative angle compared to that within in the mirage gaps, as will be discussed below.

The interference within the mirage gaps, where spin-triplet pairings dominate, is illustrated in Fig.~\ref{fig4}. 
The interference patterns exhibit a strong dependence not only on the superconducting phase difference but also on the relative orientation of the exchange fields.
The dashed curves in panels (c) and (d) of Figs.~\ref{fig3} and \ref{fig4} show scaled analytical results, with the scaling factors chosen to align their maximum values with the corresponding numerical results shown in panels (a) and (b).
The analytical and numerical results are in qualitative agreement, with the discrepancies arising from finite-size effects that are not included in the analytical treatment. 

Recent experiment has realized Andreev reflection measurements in multiterminal graphene-superconductor junction~\cite{Pandey21}.
To relate our findings to experiments, we apply the same bias voltages $V$ to the graphene electrodes, while grounding the superconducting electrodes. 
The differential conductance of the Andreev interferometer is thus given by $G(V)=(2T_{11}^A+2T_{13}^A+T_{12}+T_{14}) G_0$, where $T_{12}$ and $T_{14}$ represent the normal transmissions from electrode $1$ to the superconducting electrodes $2$ and $4$. For $|eV|<\Delta_{\rm eff}$, normal transmissions vanish at zero temperature. 
However, when $eV$ lies within the mirage gaps, the normal transmissions contribute due to the finite density of states in the mirage gaps. 
The interference patterns for differential conductance closely mirror the behavior in Figs.~\ref{fig3} and \ref{fig4} for the Andreev reflection coefficients~\cite{SM}.

{\it Conclusion.}
We have investigated the transport properties of graphene-Ising superconductor junctions. The presence of mirage gaps broadens the chemical potential range for observing SAR. Additionally, we studied the Andreev interference based on the four-terminal junction in both the main and mirage gaps. Our numerical results agree qualitatively with the analytical predictions. 
Due to the finite contribution from spin-triplet pairing, the interference can be modulated not only by the traditional method of tuning the superconducting phase difference but also by adjusting the relative orientation of the exchange fields.
Our work suggests a viable approach to control phase-coherent transport in graphene-Ising superconductor junctions. We used graphene nanoribbon as an example to illustrate the interference phenomena, however, the effects described in this work can also be realized in other materials such as bilayer graphene, which possesses inherently smaller potential fluctuations~\cite{Efetov2016}.
Our setup can be also used for Andreev interferometer-based electronic refrigerator so that the cooling power can be tuned by the relative orientation between exchange fields~\cite{Taddei25}.

\begin{acknowledgments}
{\it Acknowledgments.}
G.L. and G.T. are supported by National Natural Science Foundation of China (Grants No. 12088101 and No. 12374048) and NSAF (Grant No. U2330401). 
S.P. and W.B. acknowledge funding by the Deutsche Forschungsgemeinschaft (DFG, German Research Foundation) Project-IDs 443404566 (SPP 2244) and 467596333. 
Y.X. acknowledges support from National Natural Science Foundation of China (Grant No. 12174023).
\end{acknowledgments}

\bibliography{bib_IsingSC}{}

\begin{thebibliography}{43}%
\makeatletter
\providecommand \@ifxundefined [1]{%
 \@ifx{#1\undefined}
}%
\providecommand \@ifnum [1]{%
 \ifnum #1\expandafter \@firstoftwo
 \else \expandafter \@secondoftwo
 \fi
}%
\providecommand \@ifx [1]{%
 \ifx #1\expandafter \@firstoftwo
 \else \expandafter \@secondoftwo
 \fi
}%
\providecommand \natexlab [1]{#1}%
\providecommand \enquote  [1]{``#1''}%
\providecommand \bibnamefont  [1]{#1}%
\providecommand \bibfnamefont [1]{#1}%
\providecommand \citenamefont [1]{#1}%
\providecommand \href@noop [0]{\@secondoftwo}%
\providecommand \href [0]{\begingroup \@sanitize@url \@href}%
\providecommand \@href[1]{\@@startlink{#1}\@@href}%
\providecommand \@@href[1]{\endgroup#1\@@endlink}%
\providecommand \@sanitize@url [0]{\catcode `\\12\catcode `\$12\catcode
  `\&12\catcode `\#12\catcode `\^12\catcode `\_12\catcode `\%12\relax}%
\providecommand \@@startlink[1]{}%
\providecommand \@@endlink[0]{}%
\providecommand \url  [0]{\begingroup\@sanitize@url \@url }%
\providecommand \@url [1]{\endgroup\@href {#1}{\urlprefix }}%
\providecommand \urlprefix  [0]{URL }%
\providecommand \Eprint [0]{\href }%
\providecommand \doibase [0]{https://doi.org/}%
\providecommand \selectlanguage [0]{\@gobble}%
\providecommand \bibinfo  [0]{\@secondoftwo}%
\providecommand \bibfield  [0]{\@secondoftwo}%
\providecommand \translation [1]{[#1]}%
\providecommand \BibitemOpen [0]{}%
\providecommand \bibitemStop [0]{}%
\providecommand \bibitemNoStop [0]{.\EOS\space}%
\providecommand \EOS [0]{\spacefactor3000\relax}%
\providecommand \BibitemShut  [1]{\csname bibitem#1\endcsname}%
\let\auto@bib@innerbib\@empty
\bibitem [{\citenamefont {Ye}\ \emph {et~al.}(2012)\citenamefont {Ye},
  \citenamefont {Zhang}, \citenamefont {Akashi}, \citenamefont {Bahramy},
  \citenamefont {Arita},\ and\ \citenamefont {Iwasa}}]{Ye12}%
  \BibitemOpen
  \bibfield  {author} {\bibinfo {author} {\bibfnamefont {J.~T.}\ \bibnamefont
  {Ye}}, \bibinfo {author} {\bibfnamefont {Y.~J.}\ \bibnamefont {Zhang}},
  \bibinfo {author} {\bibfnamefont {R.}~\bibnamefont {Akashi}}, \bibinfo
  {author} {\bibfnamefont {M.~S.}\ \bibnamefont {Bahramy}}, \bibinfo {author}
  {\bibfnamefont {R.}~\bibnamefont {Arita}},\ and\ \bibinfo {author}
  {\bibfnamefont {Y.}~\bibnamefont {Iwasa}},\ }\bibfield  {title} {\bibinfo
  {title} {Superconducting dome in a gate-tuned band insulator},\ }\href
  {https://doi.org/10.1126/science.1228006} {\bibfield  {journal} {\bibinfo
  {journal} {Science}\ }\textbf {\bibinfo {volume} {338}},\ \bibinfo {pages}
  {1193} (\bibinfo {year} {2012})}\BibitemShut {NoStop}%
\bibitem [{\citenamefont {Taniguchi}\ \emph {et~al.}(2012)\citenamefont
  {Taniguchi}, \citenamefont {Matsumoto}, \citenamefont {Shimotani},\ and\
  \citenamefont {Takagi}}]{Taniguchi12}%
  \BibitemOpen
  \bibfield  {author} {\bibinfo {author} {\bibfnamefont {K.}~\bibnamefont
  {Taniguchi}}, \bibinfo {author} {\bibfnamefont {A.}~\bibnamefont
  {Matsumoto}}, \bibinfo {author} {\bibfnamefont {H.}~\bibnamefont
  {Shimotani}},\ and\ \bibinfo {author} {\bibfnamefont {H.}~\bibnamefont
  {Takagi}},\ }\bibfield  {title} {\bibinfo {title} {Electric-field-induced
  superconductivity at $9.4\,${K} in a layered transition metal disulphide
  {M}o{S}$_2$},\ }\href {https://doi.org/10.1063/1.4740268} {\bibfield
  {journal} {\bibinfo  {journal} {Appl. Phys. Lett.}\ }\textbf {\bibinfo
  {volume} {101}},\ \bibinfo {pages} {042603} (\bibinfo {year}
  {2012})}\BibitemShut {NoStop}%
\bibitem [{\citenamefont {Lu}\ \emph {et~al.}(2015)\citenamefont {Lu},
  \citenamefont {Zheliuk}, \citenamefont {Leermakers}, \citenamefont {Yuan},
  \citenamefont {Zeitler}, \citenamefont {Law},\ and\ \citenamefont
  {Ye}}]{Ye15}%
  \BibitemOpen
  \bibfield  {author} {\bibinfo {author} {\bibfnamefont {J.~M.}\ \bibnamefont
  {Lu}}, \bibinfo {author} {\bibfnamefont {O.}~\bibnamefont {Zheliuk}},
  \bibinfo {author} {\bibfnamefont {I.}~\bibnamefont {Leermakers}}, \bibinfo
  {author} {\bibfnamefont {N.~F.~Q.}\ \bibnamefont {Yuan}}, \bibinfo {author}
  {\bibfnamefont {U.}~\bibnamefont {Zeitler}}, \bibinfo {author} {\bibfnamefont
  {K.~T.}\ \bibnamefont {Law}},\ and\ \bibinfo {author} {\bibfnamefont {J.~T.}\
  \bibnamefont {Ye}},\ }\bibfield  {title} {\bibinfo {title} {Evidence for
  two-dimensional {I}sing superconductivity in gated {M}o{S}$_2$},\ }\href
  {https://doi.org/10.1126/science.aab2277} {\bibfield  {journal} {\bibinfo
  {journal} {Science}\ }\textbf {\bibinfo {volume} {350}},\ \bibinfo {pages}
  {1353} (\bibinfo {year} {2015})}\BibitemShut {NoStop}%
\bibitem [{\citenamefont {Saito}\ \emph {et~al.}(2016)\citenamefont {Saito},
  \citenamefont {Nakamura}, \citenamefont {Bahramy}, \citenamefont {Kohama},
  \citenamefont {Ye}, \citenamefont {Kasahara}, \citenamefont {Nakagawa},
  \citenamefont {Onga}, \citenamefont {Tokunaga}, \citenamefont {Nojima},
  \citenamefont {Yanase},\ and\ \citenamefont {Iwasa}}]{Saito16}%
  \BibitemOpen
  \bibfield  {author} {\bibinfo {author} {\bibfnamefont {Y.}~\bibnamefont
  {Saito}}, \bibinfo {author} {\bibfnamefont {Y.}~\bibnamefont {Nakamura}},
  \bibinfo {author} {\bibfnamefont {M.~S.}\ \bibnamefont {Bahramy}}, \bibinfo
  {author} {\bibfnamefont {Y.}~\bibnamefont {Kohama}}, \bibinfo {author}
  {\bibfnamefont {J.}~\bibnamefont {Ye}}, \bibinfo {author} {\bibfnamefont
  {Y.}~\bibnamefont {Kasahara}}, \bibinfo {author} {\bibfnamefont
  {Y.}~\bibnamefont {Nakagawa}}, \bibinfo {author} {\bibfnamefont
  {M.}~\bibnamefont {Onga}}, \bibinfo {author} {\bibfnamefont {M.}~\bibnamefont
  {Tokunaga}}, \bibinfo {author} {\bibfnamefont {T.}~\bibnamefont {Nojima}},
  \bibinfo {author} {\bibfnamefont {Y.}~\bibnamefont {Yanase}},\ and\ \bibinfo
  {author} {\bibfnamefont {Y.}~\bibnamefont {Iwasa}},\ }\bibfield  {title}
  {\bibinfo {title} {Superconductivity protected by spin--valley locking in
  ion-gated {M}o{S}$_2$},\ }\href {https://doi.org/10.1038/nphys3580}
  {\bibfield  {journal} {\bibinfo  {journal} {Nat. Phys.}\ }\textbf {\bibinfo
  {volume} {12}},\ \bibinfo {pages} {144} (\bibinfo {year} {2016})}\BibitemShut
  {NoStop}%
\bibitem [{\citenamefont {Xi}\ \emph {et~al.}(2016)\citenamefont {Xi},
  \citenamefont {Wang}, \citenamefont {Zhao}, \citenamefont {Park},
  \citenamefont {Law}, \citenamefont {Berger}, \citenamefont {Forr\'{o}},
  \citenamefont {Shan},\ and\ \citenamefont {Mak}}]{Xi16}%
  \BibitemOpen
  \bibfield  {author} {\bibinfo {author} {\bibfnamefont {X.}~\bibnamefont
  {Xi}}, \bibinfo {author} {\bibfnamefont {Z.}~\bibnamefont {Wang}}, \bibinfo
  {author} {\bibfnamefont {W.}~\bibnamefont {Zhao}}, \bibinfo {author}
  {\bibfnamefont {J.-H.}\ \bibnamefont {Park}}, \bibinfo {author}
  {\bibfnamefont {K.~T.}\ \bibnamefont {Law}}, \bibinfo {author} {\bibfnamefont
  {H.}~\bibnamefont {Berger}}, \bibinfo {author} {\bibfnamefont
  {L.}~\bibnamefont {Forr\'{o}}}, \bibinfo {author} {\bibfnamefont
  {J.}~\bibnamefont {Shan}},\ and\ \bibinfo {author} {\bibfnamefont {K.~F.}\
  \bibnamefont {Mak}},\ }\bibfield  {title} {\bibinfo {title} {Ising pairing in
  superconducting {N}b{S}e$_2$ atomic layers},\ }\href
  {https://doi.org/10.1038/nphys3538} {\bibfield  {journal} {\bibinfo
  {journal} {Nat. Phys.}\ }\textbf {\bibinfo {volume} {12}},\ \bibinfo {pages}
  {139} (\bibinfo {year} {2016})}\BibitemShut {NoStop}%
\bibitem [{\citenamefont {Tang}\ \emph
  {et~al.}(2021{\natexlab{a}})\citenamefont {Tang}, \citenamefont {Bruder},\
  and\ \citenamefont {Belzig}}]{mirage}%
  \BibitemOpen
  \bibfield  {author} {\bibinfo {author} {\bibfnamefont {G.}~\bibnamefont
  {Tang}}, \bibinfo {author} {\bibfnamefont {C.}~\bibnamefont {Bruder}},\ and\
  \bibinfo {author} {\bibfnamefont {W.}~\bibnamefont {Belzig}},\ }\bibfield
  {title} {\bibinfo {title} {Magnetic field-induced ``mirage'' gap in an
  {I}sing superconductor},\ }\href
  {https://doi.org/10.1103/PhysRevLett.126.237001} {\bibfield  {journal}
  {\bibinfo  {journal} {Phys. Rev. Lett.}\ }\textbf {\bibinfo {volume} {126}},\
  \bibinfo {pages} {237001} (\bibinfo {year} {2021}{\natexlab{a}})}\BibitemShut
  {NoStop}%
\bibitem [{\citenamefont {Patil}\ \emph {et~al.}(2023)\citenamefont {Patil},
  \citenamefont {Tang},\ and\ \citenamefont {Belzig}}]{ising_patil}%
  \BibitemOpen
  \bibfield  {author} {\bibinfo {author} {\bibfnamefont {S.}~\bibnamefont
  {Patil}}, \bibinfo {author} {\bibfnamefont {G.}~\bibnamefont {Tang}},\ and\
  \bibinfo {author} {\bibfnamefont {W.}~\bibnamefont {Belzig}},\ }\bibfield
  {title} {\bibinfo {title} {Spectral properties of a mixed singlet-triplet
  {I}sing superconductor},\ }\href
  {https://doi.org/https://doi.org/10.3389/femat.2023.1254302} {\bibfield
  {journal} {\bibinfo  {journal} {Front. Electron. Mater.}\ }\textbf {\bibinfo
  {volume} {3}},\ \bibinfo {pages} {1254302} (\bibinfo {year}
  {2023})}\BibitemShut {NoStop}%
\bibitem [{\citenamefont {Ili\ifmmode~\acute{c}\else \'{c}\fi{}}\ \emph
  {et~al.}(2023)\citenamefont {Ili\ifmmode~\acute{c}\else \'{c}\fi{}},
  \citenamefont {Meyer},\ and\ \citenamefont {Houzet}}]{ilic_mirage_2023}%
  \BibitemOpen
  \bibfield  {author} {\bibinfo {author} {\bibfnamefont {S.}~\bibnamefont
  {Ili\ifmmode~\acute{c}\else \'{c}\fi{}}}, \bibinfo {author} {\bibfnamefont
  {J.~S.}\ \bibnamefont {Meyer}},\ and\ \bibinfo {author} {\bibfnamefont
  {M.}~\bibnamefont {Houzet}},\ }\bibfield  {title} {\bibinfo {title} {Spectral
  properties of disordered {I}sing superconductors with singlet and triplet
  pairing in in-plane magnetic fields},\ }\href
  {https://doi.org/10.1103/PhysRevB.108.214510} {\bibfield  {journal} {\bibinfo
   {journal} {Phys. Rev. B}\ }\textbf {\bibinfo {volume} {108}},\ \bibinfo
  {pages} {214510} (\bibinfo {year} {2023})}\BibitemShut {NoStop}%
\bibitem [{\citenamefont {M\"ockli}\ and\ \citenamefont
  {Khodas}(2019)}]{Mockli19}%
  \BibitemOpen
  \bibfield  {author} {\bibinfo {author} {\bibfnamefont {D.}~\bibnamefont
  {M\"ockli}}\ and\ \bibinfo {author} {\bibfnamefont {M.}~\bibnamefont
  {Khodas}},\ }\bibfield  {title} {\bibinfo {title} {Magnetic-field induced
  $s+\mathit{if}$ pairing in {I}sing superconductors},\ }\href
  {https://doi.org/10.1103/PhysRevB.99.180505} {\bibfield  {journal} {\bibinfo
  {journal} {Phys. Rev. B}\ }\textbf {\bibinfo {volume} {99}},\ \bibinfo
  {pages} {180505} (\bibinfo {year} {2019})}\BibitemShut {NoStop}%
\bibitem [{\citenamefont {M\"ockli}\ and\ \citenamefont
  {Khodas}(2020)}]{Mockli20}%
  \BibitemOpen
  \bibfield  {author} {\bibinfo {author} {\bibfnamefont {D.}~\bibnamefont
  {M\"ockli}}\ and\ \bibinfo {author} {\bibfnamefont {M.}~\bibnamefont
  {Khodas}},\ }\bibfield  {title} {\bibinfo {title} {Ising superconductors:
  Interplay of magnetic field, triplet channels, and disorder},\ }\href
  {https://doi.org/10.1103/PhysRevB.101.014510} {\bibfield  {journal} {\bibinfo
   {journal} {Phys. Rev. B}\ }\textbf {\bibinfo {volume} {101}},\ \bibinfo
  {pages} {014510} (\bibinfo {year} {2020})}\BibitemShut {NoStop}%
\bibitem [{\citenamefont {Wickramaratne}\ \emph {et~al.}(2020)\citenamefont
  {Wickramaratne}, \citenamefont {Khmelevskyi}, \citenamefont {Agterberg},\
  and\ \citenamefont {Mazin}}]{Wickramaratne20}%
  \BibitemOpen
  \bibfield  {author} {\bibinfo {author} {\bibfnamefont {D.}~\bibnamefont
  {Wickramaratne}}, \bibinfo {author} {\bibfnamefont {S.}~\bibnamefont
  {Khmelevskyi}}, \bibinfo {author} {\bibfnamefont {D.~F.}\ \bibnamefont
  {Agterberg}},\ and\ \bibinfo {author} {\bibfnamefont {I.~I.}\ \bibnamefont
  {Mazin}},\ }\bibfield  {title} {\bibinfo {title} {Ising superconductivity and
  magnetism in {N}b{S}e$_2$},\ }\href
  {https://doi.org/10.1103/PhysRevX.10.041003} {\bibfield  {journal} {\bibinfo
  {journal} {Phys. Rev. X}\ }\textbf {\bibinfo {volume} {10}},\ \bibinfo
  {pages} {041003} (\bibinfo {year} {2020})}\BibitemShut {NoStop}%
\bibitem [{\citenamefont {Zhou}\ \emph {et~al.}(2016)\citenamefont {Zhou},
  \citenamefont {Yuan}, \citenamefont {Jiang},\ and\ \citenamefont
  {Law}}]{Zhou16}%
  \BibitemOpen
  \bibfield  {author} {\bibinfo {author} {\bibfnamefont {B.~T.}\ \bibnamefont
  {Zhou}}, \bibinfo {author} {\bibfnamefont {N.~F.~Q.}\ \bibnamefont {Yuan}},
  \bibinfo {author} {\bibfnamefont {H.-L.}\ \bibnamefont {Jiang}},\ and\
  \bibinfo {author} {\bibfnamefont {K.~T.}\ \bibnamefont {Law}},\ }\bibfield
  {title} {\bibinfo {title} {Ising superconductivity and {M}ajorana fermions in
  transition-metal dichalcogenides},\ }\href
  {https://doi.org/10.1103/PhysRevB.93.180501} {\bibfield  {journal} {\bibinfo
  {journal} {Phys. Rev. B}\ }\textbf {\bibinfo {volume} {93}},\ \bibinfo
  {pages} {180501(R)} (\bibinfo {year} {2016})}\BibitemShut {NoStop}%
\bibitem [{\citenamefont {Lv}\ \emph {et~al.}(2018)\citenamefont {Lv},
  \citenamefont {Zhou}, \citenamefont {Yang},\ and\ \citenamefont
  {Sun}}]{Transport_Sun18}%
  \BibitemOpen
  \bibfield  {author} {\bibinfo {author} {\bibfnamefont {P.}~\bibnamefont
  {Lv}}, \bibinfo {author} {\bibfnamefont {Y.-F.}\ \bibnamefont {Zhou}},
  \bibinfo {author} {\bibfnamefont {N.-X.}\ \bibnamefont {Yang}},\ and\
  \bibinfo {author} {\bibfnamefont {Q.-F.}\ \bibnamefont {Sun}},\ }\bibfield
  {title} {\bibinfo {title} {Magnetoanisotropic spin-triplet {A}ndreev
  reflection in ferromagnet-{I}sing superconductor junctions},\ }\href
  {https://doi.org/10.1103/PhysRevB.97.144501} {\bibfield  {journal} {\bibinfo
  {journal} {Phys. Rev. B}\ }\textbf {\bibinfo {volume} {97}},\ \bibinfo
  {pages} {144501} (\bibinfo {year} {2018})}\BibitemShut {NoStop}%
\bibitem [{\citenamefont {Cheng}\ and\ \citenamefont
  {Sun}(2019)}]{Transport_Sun19}%
  \BibitemOpen
  \bibfield  {author} {\bibinfo {author} {\bibfnamefont {Q.}~\bibnamefont
  {Cheng}}\ and\ \bibinfo {author} {\bibfnamefont {Q.-F.}\ \bibnamefont
  {Sun}},\ }\bibfield  {title} {\bibinfo {title} {Switch effect and
  0-$\ensuremath{\pi}$ transition in {I}sing superconductor {J}osephson
  junctions},\ }\href {https://doi.org/10.1103/PhysRevB.99.184507} {\bibfield
  {journal} {\bibinfo  {journal} {Phys. Rev. B}\ }\textbf {\bibinfo {volume}
  {99}},\ \bibinfo {pages} {184507} (\bibinfo {year} {2019})}\BibitemShut
  {NoStop}%
\bibitem [{\citenamefont {Tang}\ \emph
  {et~al.}(2021{\natexlab{b}})\citenamefont {Tang}, \citenamefont {Klees},
  \citenamefont {Bruder},\ and\ \citenamefont {Belzig}}]{GT21}%
  \BibitemOpen
  \bibfield  {author} {\bibinfo {author} {\bibfnamefont {G.}~\bibnamefont
  {Tang}}, \bibinfo {author} {\bibfnamefont {R.~L.}\ \bibnamefont {Klees}},
  \bibinfo {author} {\bibfnamefont {C.}~\bibnamefont {Bruder}},\ and\ \bibinfo
  {author} {\bibfnamefont {W.}~\bibnamefont {Belzig}},\ }\bibfield  {title}
  {\bibinfo {title} {Controlling charge and spin transport in an
  {I}sing-superconductor {J}osephson junction},\ }\href
  {https://doi.org/10.1103/PhysRevB.104.L241413} {\bibfield  {journal}
  {\bibinfo  {journal} {Phys. Rev. B}\ }\textbf {\bibinfo {volume} {104}},\
  \bibinfo {pages} {L241413} (\bibinfo {year}
  {2021}{\natexlab{b}})}\BibitemShut {NoStop}%
\bibitem [{\citenamefont {Lu}\ \emph {et~al.}(2022)\citenamefont {Lu},
  \citenamefont {Sun},\ and\ \citenamefont {Cheng}}]{Transport_Sun22_1}%
  \BibitemOpen
  \bibfield  {author} {\bibinfo {author} {\bibfnamefont {W.-T.}\ \bibnamefont
  {Lu}}, \bibinfo {author} {\bibfnamefont {Q.-F.}\ \bibnamefont {Sun}},\ and\
  \bibinfo {author} {\bibfnamefont {Q.}~\bibnamefont {Cheng}},\ }\bibfield
  {title} {\bibinfo {title} {Equal-spin and oblique-spin crossed {A}ndreev
  reflections in ferromagnet/{I}sing superconductor/ferromagnet junction},\
  }\href {https://doi.org/10.1103/PhysRevB.105.125425} {\bibfield  {journal}
  {\bibinfo  {journal} {Phys. Rev. B}\ }\textbf {\bibinfo {volume} {105}},\
  \bibinfo {pages} {125425} (\bibinfo {year} {2022})}\BibitemShut {NoStop}%
\bibitem [{\citenamefont {Dai}\ \emph {et~al.}(2022)\citenamefont {Dai},
  \citenamefont {Mao},\ and\ \citenamefont {Sun}}]{Transport_Sun22_2}%
  \BibitemOpen
  \bibfield  {author} {\bibinfo {author} {\bibfnamefont {Y.-X.}\ \bibnamefont
  {Dai}}, \bibinfo {author} {\bibfnamefont {Y.}~\bibnamefont {Mao}},\ and\
  \bibinfo {author} {\bibfnamefont {Q.-F.}\ \bibnamefont {Sun}},\ }\bibfield
  {title} {\bibinfo {title} {Spin transport in a normal metal--{I}sing
  superconductor junction},\ }\href
  {https://doi.org/10.1103/PhysRevB.106.184513} {\bibfield  {journal} {\bibinfo
   {journal} {Phys. Rev. B}\ }\textbf {\bibinfo {volume} {106}},\ \bibinfo
  {pages} {184513} (\bibinfo {year} {2022})}\BibitemShut {NoStop}%
\bibitem [{\citenamefont {Asano}\ and\ \citenamefont {Yanase}(2024)}]{Asano24}%
  \BibitemOpen
  \bibfield  {author} {\bibinfo {author} {\bibfnamefont {S.}~\bibnamefont
  {Asano}}\ and\ \bibinfo {author} {\bibfnamefont {Y.}~\bibnamefont {Yanase}},\
  }\bibfield  {title} {\bibinfo {title} {Tuning monolayer superconductivity in
  twisted {N}b{S}e$_2$ graphene heterostructures},\ }\href
  {https://doi.org/10.1103/PhysRevB.110.134516} {\bibfield  {journal} {\bibinfo
   {journal} {Phys. Rev. B}\ }\textbf {\bibinfo {volume} {110}},\ \bibinfo
  {pages} {134516} (\bibinfo {year} {2024})}\BibitemShut {NoStop}%
\bibitem [{\citenamefont {Li}\ \emph {et~al.}(2024)\citenamefont {Li},
  \citenamefont {Qi}, \citenamefont {Wu},\ and\ \citenamefont {He}}]{He24}%
  \BibitemOpen
  \bibfield  {author} {\bibinfo {author} {\bibfnamefont {X.-Z.}\ \bibnamefont
  {Li}}, \bibinfo {author} {\bibfnamefont {Z.-B.}\ \bibnamefont {Qi}}, \bibinfo
  {author} {\bibfnamefont {Q.}~\bibnamefont {Wu}},\ and\ \bibinfo {author}
  {\bibfnamefont {W.-Y.}\ \bibnamefont {He}},\ }\bibfield  {title} {\bibinfo
  {title} {Topological superconductivity in monolayer
  {T}$_{\textrm{d}}$-{M}o{T}e$_2$},\ }\href
  {https://doi.org/10.1038/s42005-024-01881-6} {\bibfield  {journal} {\bibinfo
  {journal} {Commun. Phys.}\ }\textbf {\bibinfo {volume} {7}},\ \bibinfo
  {pages} {396} (\bibinfo {year} {2024})}\BibitemShut {NoStop}%
\bibitem [{\citenamefont {Patil}\ \emph {et~al.}(2025)\citenamefont {Patil},
  \citenamefont {Tang},\ and\ \citenamefont {Belzig}}]{GT24}%
  \BibitemOpen
  \bibfield  {author} {\bibinfo {author} {\bibfnamefont {S.}~\bibnamefont
  {Patil}}, \bibinfo {author} {\bibfnamefont {G.}~\bibnamefont {Tang}},\ and\
  \bibinfo {author} {\bibfnamefont {W.}~\bibnamefont {Belzig}},\ }\bibfield
  {title} {\bibinfo {title} {Spin-split {A}ndreev bound states and diode effect
  in an {I}sing superconductor {J}osephson junction},\ }\href
  {https://doi.org/10.1103/PhysRevB.111.L060502} {\bibfield  {journal}
  {\bibinfo  {journal} {Phys. Rev. B}\ }\textbf {\bibinfo {volume} {111}},\
  \bibinfo {pages} {L060502} (\bibinfo {year} {2025})}\BibitemShut {NoStop}%
\bibitem [{\citenamefont {Idzuchi}\ \emph {et~al.}(2021)\citenamefont
  {Idzuchi}, \citenamefont {Pientka}, \citenamefont {Huang}, \citenamefont
  {Harada}, \citenamefont {G\"{u}l}, \citenamefont {Shin}, \citenamefont
  {Nguyen}, \citenamefont {Jo}, \citenamefont {Shindo}, \citenamefont {Cava},
  \citenamefont {Canfield},\ and\ \citenamefont {Kim}}]{Idzuchi21}%
  \BibitemOpen
  \bibfield  {author} {\bibinfo {author} {\bibfnamefont {H.}~\bibnamefont
  {Idzuchi}}, \bibinfo {author} {\bibfnamefont {F.}~\bibnamefont {Pientka}},
  \bibinfo {author} {\bibfnamefont {K.-F.}\ \bibnamefont {Huang}}, \bibinfo
  {author} {\bibfnamefont {K.}~\bibnamefont {Harada}}, \bibinfo {author}
  {\bibfnamefont {O.}~\bibnamefont {G\"{u}l}}, \bibinfo {author} {\bibfnamefont
  {Y.~J.}\ \bibnamefont {Shin}}, \bibinfo {author} {\bibfnamefont {L.~T.}\
  \bibnamefont {Nguyen}}, \bibinfo {author} {\bibfnamefont {N.~H.}\
  \bibnamefont {Jo}}, \bibinfo {author} {\bibfnamefont {D.}~\bibnamefont
  {Shindo}}, \bibinfo {author} {\bibfnamefont {R.~J.}\ \bibnamefont {Cava}},
  \bibinfo {author} {\bibfnamefont {P.~C.}\ \bibnamefont {Canfield}},\ and\
  \bibinfo {author} {\bibfnamefont {P.}~\bibnamefont {Kim}},\ }\bibfield
  {title} {\bibinfo {title} {Unconventional supercurrent phase in {I}sing
  superconductor {J}osephson junction with atomically thin magnetic
  insulator},\ }\href
  {https://doi.org/https://doi.org/10.1038/s41467-021-25608-1} {\bibfield
  {journal} {\bibinfo  {journal} {Nat. Commun.}\ }\textbf {\bibinfo {volume}
  {12}},\ \bibinfo {pages} {1} (\bibinfo {year} {2021})}\BibitemShut {NoStop}%
\bibitem [{\citenamefont {Jeon}\ \emph {et~al.}(2021)\citenamefont {Jeon},
  \citenamefont {Cho}, \citenamefont {Chakraborty}, \citenamefont {Jeon},
  \citenamefont {Yoon}, \citenamefont {Han}, \citenamefont {Kim},\ and\
  \citenamefont {Parkin}}]{Jeon21}%
  \BibitemOpen
  \bibfield  {author} {\bibinfo {author} {\bibfnamefont {K.-R.}\ \bibnamefont
  {Jeon}}, \bibinfo {author} {\bibfnamefont {K.}~\bibnamefont {Cho}}, \bibinfo
  {author} {\bibfnamefont {A.}~\bibnamefont {Chakraborty}}, \bibinfo {author}
  {\bibfnamefont {J.-C.}\ \bibnamefont {Jeon}}, \bibinfo {author}
  {\bibfnamefont {J.}~\bibnamefont {Yoon}}, \bibinfo {author} {\bibfnamefont
  {H.}~\bibnamefont {Han}}, \bibinfo {author} {\bibfnamefont {J.-K.}\
  \bibnamefont {Kim}},\ and\ \bibinfo {author} {\bibfnamefont {S.~S.~P.}\
  \bibnamefont {Parkin}},\ }\bibfield  {title} {\bibinfo {title} {Role of
  two-dimensional {I}sing superconductivity in the nonequilibrium quasiparticle
  spin-to-charge conversion efficiency},\ }\href
  {https://doi.org/10.1021/acsnano.1c07192} {\bibfield  {journal} {\bibinfo
  {journal} {ACS Nano}\ }\textbf {\bibinfo {volume} {15}},\ \bibinfo {pages}
  {16819} (\bibinfo {year} {2021})}\BibitemShut {NoStop}%
\bibitem [{\citenamefont {Kang}\ \emph {et~al.}(2022)\citenamefont {Kang},
  \citenamefont {Berger}, \citenamefont {Watanabe}, \citenamefont {Taniguchi},
  \citenamefont {Forr\'{o}}, \citenamefont {Shan},\ and\ \citenamefont
  {Mak}}]{Kang22}%
  \BibitemOpen
  \bibfield  {author} {\bibinfo {author} {\bibfnamefont {K.}~\bibnamefont
  {Kang}}, \bibinfo {author} {\bibfnamefont {H.}~\bibnamefont {Berger}},
  \bibinfo {author} {\bibfnamefont {K.}~\bibnamefont {Watanabe}}, \bibinfo
  {author} {\bibfnamefont {T.}~\bibnamefont {Taniguchi}}, \bibinfo {author}
  {\bibfnamefont {L.}~\bibnamefont {Forr\'{o}}}, \bibinfo {author}
  {\bibfnamefont {J.}~\bibnamefont {Shan}},\ and\ \bibinfo {author}
  {\bibfnamefont {K.~F.}\ \bibnamefont {Mak}},\ }\bibfield  {title} {\bibinfo
  {title} {van der {W}aals $\pi$ {J}osephson junctions},\ }\href
  {https://doi.org/10.1021/acs.nanolett.2c01640} {\bibfield  {journal}
  {\bibinfo  {journal} {Nano Lett.}\ }\textbf {\bibinfo {volume} {22}},\
  \bibinfo {pages} {5510} (\bibinfo {year} {2022})}\BibitemShut {NoStop}%
\bibitem [{\citenamefont {Zalic}\ \emph {et~al.}(2023)\citenamefont {Zalic},
  \citenamefont {Taniguchi}, \citenamefont {Watanabe}, \citenamefont {Gazit},\
  and\ \citenamefont {Steinberg}}]{Zalic23}%
  \BibitemOpen
  \bibfield  {author} {\bibinfo {author} {\bibfnamefont {A.}~\bibnamefont
  {Zalic}}, \bibinfo {author} {\bibfnamefont {T.}~\bibnamefont {Taniguchi}},
  \bibinfo {author} {\bibfnamefont {K.}~\bibnamefont {Watanabe}}, \bibinfo
  {author} {\bibfnamefont {S.}~\bibnamefont {Gazit}},\ and\ \bibinfo {author}
  {\bibfnamefont {H.}~\bibnamefont {Steinberg}},\ }\bibfield  {title} {\bibinfo
  {title} {High magnetic field stability in a planar graphene-{N}b{S}e$_2$
  {SQUID}},\ }\href {https://doi.org/10.1021/acs.nanolett.3c01552} {\bibfield
  {journal} {\bibinfo  {journal} {Nano Lett.}\ }\textbf {\bibinfo {volume}
  {23}},\ \bibinfo {pages} {6102} (\bibinfo {year} {2023})}\BibitemShut
  {NoStop}%
\bibitem [{\citenamefont {Xiong}\ \emph {et~al.}(2024)\citenamefont {Xiong},
  \citenamefont {Xie}, \citenamefont {Cheng}, \citenamefont {Dai},
  \citenamefont {Cui}, \citenamefont {Wang}, \citenamefont {Liu}, \citenamefont
  {Zhou}, \citenamefont {Wang}, \citenamefont {Xu}, \citenamefont {Chen},
  \citenamefont {Cheong}, \citenamefont {Liang},\ and\ \citenamefont
  {Miao}}]{Xiong24}%
  \BibitemOpen
  \bibfield  {author} {\bibinfo {author} {\bibfnamefont {J.}~\bibnamefont
  {Xiong}}, \bibinfo {author} {\bibfnamefont {J.}~\bibnamefont {Xie}}, \bibinfo
  {author} {\bibfnamefont {B.}~\bibnamefont {Cheng}}, \bibinfo {author}
  {\bibfnamefont {Y.}~\bibnamefont {Dai}}, \bibinfo {author} {\bibfnamefont
  {X.}~\bibnamefont {Cui}}, \bibinfo {author} {\bibfnamefont {L.}~\bibnamefont
  {Wang}}, \bibinfo {author} {\bibfnamefont {Z.}~\bibnamefont {Liu}}, \bibinfo
  {author} {\bibfnamefont {J.}~\bibnamefont {Zhou}}, \bibinfo {author}
  {\bibfnamefont {N.}~\bibnamefont {Wang}}, \bibinfo {author} {\bibfnamefont
  {X.}~\bibnamefont {Xu}}, \bibinfo {author} {\bibfnamefont {X.}~\bibnamefont
  {Chen}}, \bibinfo {author} {\bibfnamefont {S.-W.}\ \bibnamefont {Cheong}},
  \bibinfo {author} {\bibfnamefont {S.-J.}\ \bibnamefont {Liang}},\ and\
  \bibinfo {author} {\bibfnamefont {F.}~\bibnamefont {Miao}},\ }\bibfield
  {title} {\bibinfo {title} {Electrical switching of {I}sing-superconducting
  nonreciprocity for quantum neuronal transistor},\ }\href
  {https://doi.org/10.1038/s41467-024-48882-1} {\bibfield  {journal} {\bibinfo
  {journal} {Nat. Commun.}\ }\textbf {\bibinfo {volume} {15}},\ \bibinfo
  {pages} {4953} (\bibinfo {year} {2024})}\BibitemShut {NoStop}%
\bibitem [{\citenamefont {Beenakker}(2006)}]{Specular_06}%
  \BibitemOpen
  \bibfield  {author} {\bibinfo {author} {\bibfnamefont {C.~W.~J.}\
  \bibnamefont {Beenakker}},\ }\bibfield  {title} {\bibinfo {title} {Specular
  {A}ndreev reflection in graphene},\ }\href
  {https://doi.org/10.1103/PhysRevLett.97.067007} {\bibfield  {journal}
  {\bibinfo  {journal} {Phys. Rev. Lett.}\ }\textbf {\bibinfo {volume} {97}},\
  \bibinfo {pages} {067007} (\bibinfo {year} {2006})}\BibitemShut {NoStop}%
\bibitem [{\citenamefont {Ram}\ \emph {et~al.}(2023)\citenamefont {Ram},
  \citenamefont {Beckmann}, \citenamefont {Danneau},\ and\ \citenamefont
  {Belzig}}]{Ram23}%
  \BibitemOpen
  \bibfield  {author} {\bibinfo {author} {\bibfnamefont {P.}~\bibnamefont
  {Ram}}, \bibinfo {author} {\bibfnamefont {D.}~\bibnamefont {Beckmann}},
  \bibinfo {author} {\bibfnamefont {R.}~\bibnamefont {Danneau}},\ and\ \bibinfo
  {author} {\bibfnamefont {W.}~\bibnamefont {Belzig}},\ }\bibfield  {title}
  {\bibinfo {title} {Andreev and normal reflections in gapped bilayer
  graphene--superconductor junctions},\ }\href
  {https://doi.org/10.1103/PhysRevB.108.184510} {\bibfield  {journal} {\bibinfo
   {journal} {Phys. Rev. B}\ }\textbf {\bibinfo {volume} {108}},\ \bibinfo
  {pages} {184510} (\bibinfo {year} {2023})}\BibitemShut {NoStop}%
\bibitem [{\citenamefont {Efetov}\ \emph {et~al.}(2016)\citenamefont {Efetov},
  \citenamefont {Wang}, \citenamefont {Handschin}, \citenamefont {Efetov},
  \citenamefont {Shuang}, \citenamefont {Cava}, \citenamefont {Taniguchi},
  \citenamefont {Watanabe}, \citenamefont {Hone}, \citenamefont {Dean},\ and\
  \citenamefont {Kim}}]{Efetov2016}%
  \BibitemOpen
  \bibfield  {author} {\bibinfo {author} {\bibfnamefont {D.~K.}\ \bibnamefont
  {Efetov}}, \bibinfo {author} {\bibfnamefont {L.}~\bibnamefont {Wang}},
  \bibinfo {author} {\bibfnamefont {C.}~\bibnamefont {Handschin}}, \bibinfo
  {author} {\bibfnamefont {K.~B.}\ \bibnamefont {Efetov}}, \bibinfo {author}
  {\bibfnamefont {J.}~\bibnamefont {Shuang}}, \bibinfo {author} {\bibfnamefont
  {R.}~\bibnamefont {Cava}}, \bibinfo {author} {\bibfnamefont {T.}~\bibnamefont
  {Taniguchi}}, \bibinfo {author} {\bibfnamefont {K.}~\bibnamefont {Watanabe}},
  \bibinfo {author} {\bibfnamefont {J.}~\bibnamefont {Hone}}, \bibinfo {author}
  {\bibfnamefont {C.~R.}\ \bibnamefont {Dean}},\ and\ \bibinfo {author}
  {\bibfnamefont {P.}~\bibnamefont {Kim}},\ }\bibfield  {title} {\bibinfo
  {title} {Specular interband {A}ndreev reflections at van der {W}aals
  interfaces between graphene and {N}b{S}e$_2$},\ }\href
  {https://doi.org/10.1038/nphys3583} {\bibfield  {journal} {\bibinfo
  {journal} {Nat. Phys.}\ }\textbf {\bibinfo {volume} {12}},\ \bibinfo {pages}
  {328} (\bibinfo {year} {2016})}\BibitemShut {NoStop}%
\bibitem [{\citenamefont {Efetov}\ and\ \citenamefont
  {Efetov}(2016)}]{Efetov16}%
  \BibitemOpen
  \bibfield  {author} {\bibinfo {author} {\bibfnamefont {D.~K.}\ \bibnamefont
  {Efetov}}\ and\ \bibinfo {author} {\bibfnamefont {K.~B.}\ \bibnamefont
  {Efetov}},\ }\bibfield  {title} {\bibinfo {title} {Crossover from retro to
  specular {A}ndreev reflections in bilayer graphene},\ }\href
  {https://doi.org/10.1103/PhysRevB.94.075403} {\bibfield  {journal} {\bibinfo
  {journal} {Phys. Rev. B}\ }\textbf {\bibinfo {volume} {94}},\ \bibinfo
  {pages} {075403} (\bibinfo {year} {2016})}\BibitemShut {NoStop}%
\bibitem [{\citenamefont {Ludwig}(2007)}]{Ludwig07}%
  \BibitemOpen
  \bibfield  {author} {\bibinfo {author} {\bibfnamefont {T.}~\bibnamefont
  {Ludwig}},\ }\bibfield  {title} {\bibinfo {title} {Andreev reflection in
  bilayer graphene},\ }\href {https://doi.org/10.1103/PhysRevB.75.195322}
  {\bibfield  {journal} {\bibinfo  {journal} {Phys. Rev. B}\ }\textbf {\bibinfo
  {volume} {75}},\ \bibinfo {pages} {195322} (\bibinfo {year}
  {2007})}\BibitemShut {NoStop}%
\bibitem [{\citenamefont {Takane}\ \emph {et~al.}(2017)\citenamefont {Takane},
  \citenamefont {Yarimizu},\ and\ \citenamefont {Kanda}}]{Takane17}%
  \BibitemOpen
  \bibfield  {author} {\bibinfo {author} {\bibfnamefont {Y.}~\bibnamefont
  {Takane}}, \bibinfo {author} {\bibfnamefont {K.}~\bibnamefont {Yarimizu}},\
  and\ \bibinfo {author} {\bibfnamefont {A.}~\bibnamefont {Kanda}},\ }\bibfield
   {title} {\bibinfo {title} {Andreev reflection in a bilayer graphene
  junction: Role of spatial variation of the charge neutrality point},\ }\href
  {https://doi.org/10.7566/JPSJ.86.064707} {\bibfield  {journal} {\bibinfo
  {journal} {J. Phys. Soc. Jpn.}\ }\textbf {\bibinfo {volume} {86}},\ \bibinfo
  {pages} {064707} (\bibinfo {year} {2017})}\BibitemShut {NoStop}%
\bibitem [{\citenamefont {Cheng}\ \emph {et~al.}(2009)\citenamefont {Cheng},
  \citenamefont {Xing}, \citenamefont {Wang},\ and\ \citenamefont
  {Sun}}]{Specular_Sun_09}%
  \BibitemOpen
  \bibfield  {author} {\bibinfo {author} {\bibfnamefont {S.-g.}\ \bibnamefont
  {Cheng}}, \bibinfo {author} {\bibfnamefont {Y.}~\bibnamefont {Xing}},
  \bibinfo {author} {\bibfnamefont {J.}~\bibnamefont {Wang}},\ and\ \bibinfo
  {author} {\bibfnamefont {Q.-f.}\ \bibnamefont {Sun}},\ }\bibfield  {title}
  {\bibinfo {title} {Controllable {A}ndreev retroreflection and specular
  {A}ndreev reflection in a four-terminal graphene-superconductor hybrid
  system},\ }\href {https://doi.org/10.1103/PhysRevLett.103.167003} {\bibfield
  {journal} {\bibinfo  {journal} {Phys. Rev. Lett.}\ }\textbf {\bibinfo
  {volume} {103}},\ \bibinfo {pages} {167003} (\bibinfo {year}
  {2009})}\BibitemShut {NoStop}%
\bibitem [{\citenamefont {Liu}\ \emph {et~al.}(2024)\citenamefont {Liu},
  \citenamefont {Mao},\ and\ \citenamefont {Sun}}]{Sun24}%
  \BibitemOpen
  \bibfield  {author} {\bibinfo {author} {\bibfnamefont {P.-Y.}\ \bibnamefont
  {Liu}}, \bibinfo {author} {\bibfnamefont {Y.}~\bibnamefont {Mao}},\ and\
  \bibinfo {author} {\bibfnamefont {Q.-F.}\ \bibnamefont {Sun}},\ }\bibfield
  {title} {\bibinfo {title} {Four-terminal graphene-superconductor thermal
  switch controlled by the superconducting phase difference},\ }\href
  {https://doi.org/10.1103/PhysRevApplied.21.024001} {\bibfield  {journal}
  {\bibinfo  {journal} {Phys. Rev. Appl.}\ }\textbf {\bibinfo {volume} {21}},\
  \bibinfo {pages} {024001} (\bibinfo {year} {2024})}\BibitemShut {NoStop}%
\bibitem [{\citenamefont {Sancho}\ \emph {et~al.}(1984)\citenamefont {Sancho},
  \citenamefont {Sancho},\ and\ \citenamefont {Rubio}}]{TransferMatrix1}%
  \BibitemOpen
  \bibfield  {author} {\bibinfo {author} {\bibfnamefont {M.~P.~L.}\
  \bibnamefont {Sancho}}, \bibinfo {author} {\bibfnamefont {J.~M.~L.}\
  \bibnamefont {Sancho}},\ and\ \bibinfo {author} {\bibfnamefont
  {J.}~\bibnamefont {Rubio}},\ }\bibfield  {title} {\bibinfo {title} {Quick
  iterative scheme for the calculation of transfer matrices: application to
  {M}o (100)},\ }\href {https://doi.org/10.1088/0305-4608/14/5/016} {\bibfield
  {journal} {\bibinfo  {journal} {J. Phys. F: Met. Phys.}\ }\textbf {\bibinfo
  {volume} {14}},\ \bibinfo {pages} {1205} (\bibinfo {year}
  {1984})}\BibitemShut {NoStop}%
\bibitem [{\citenamefont {Sancho}\ \emph {et~al.}(1985)\citenamefont {Sancho},
  \citenamefont {Sancho}, \citenamefont {Sancho},\ and\ \citenamefont
  {Rubio}}]{TransferMatrix2}%
  \BibitemOpen
  \bibfield  {author} {\bibinfo {author} {\bibfnamefont {M.~P.~L.}\
  \bibnamefont {Sancho}}, \bibinfo {author} {\bibfnamefont {J.~M.~L.}\
  \bibnamefont {Sancho}}, \bibinfo {author} {\bibfnamefont {J.~M.~L.}\
  \bibnamefont {Sancho}},\ and\ \bibinfo {author} {\bibfnamefont
  {J.}~\bibnamefont {Rubio}},\ }\bibfield  {title} {\bibinfo {title} {Highly
  convergent schemes for the calculation of bulk and surface {G}reen
  functions},\ }\href {https://doi.org/10.1088/0305-4608/15/4/009} {\bibfield
  {journal} {\bibinfo  {journal} {J. Phys. F: Met. Phys.}\ }\textbf {\bibinfo
  {volume} {15}},\ \bibinfo {pages} {851} (\bibinfo {year} {1985})}\BibitemShut
  {NoStop}%
\bibitem [{SM()}]{SM}%
  \BibitemOpen
  \href@noop {} {\bibinfo {title} {{See Supplemental Material which includes
  Ref.~\cite{noneqSC} for details.}}}\BibitemShut {Stop}%
\bibitem [{\citenamefont {Xing}\ \emph {et~al.}(2011)\citenamefont {Xing},
  \citenamefont {Wang},\ and\ \citenamefont {Sun}}]{Specular_Sun_11}%
  \BibitemOpen
  \bibfield  {author} {\bibinfo {author} {\bibfnamefont {Y.}~\bibnamefont
  {Xing}}, \bibinfo {author} {\bibfnamefont {J.}~\bibnamefont {Wang}},\ and\
  \bibinfo {author} {\bibfnamefont {Q.-f.}\ \bibnamefont {Sun}},\ }\bibfield
  {title} {\bibinfo {title} {Parity of specular {A}ndreev reflection under a
  mirror operation in a zigzag graphene ribbon},\ }\href
  {https://doi.org/10.1103/PhysRevB.83.205418} {\bibfield  {journal} {\bibinfo
  {journal} {Phys. Rev. B}\ }\textbf {\bibinfo {volume} {83}},\ \bibinfo
  {pages} {205418} (\bibinfo {year} {2011})}\BibitemShut {NoStop}%
\bibitem [{\citenamefont {Nakabayashi}\ \emph {et~al.}(2009)\citenamefont
  {Nakabayashi}, \citenamefont {Yamamoto},\ and\ \citenamefont
  {Kurihara}}]{Parity_odd_09}%
  \BibitemOpen
  \bibfield  {author} {\bibinfo {author} {\bibfnamefont {J.}~\bibnamefont
  {Nakabayashi}}, \bibinfo {author} {\bibfnamefont {D.}~\bibnamefont
  {Yamamoto}},\ and\ \bibinfo {author} {\bibfnamefont {S.}~\bibnamefont
  {Kurihara}},\ }\bibfield  {title} {\bibinfo {title} {Band-selective filter in
  a zigzag graphene nanoribbon},\ }\href
  {https://doi.org/10.1103/PhysRevLett.102.066803} {\bibfield  {journal}
  {\bibinfo  {journal} {Phys. Rev. Lett.}\ }\textbf {\bibinfo {volume} {102}},\
  \bibinfo {pages} {066803} (\bibinfo {year} {2009})}\BibitemShut {NoStop}%
\bibitem [{\citenamefont {Akhmerov}\ \emph {et~al.}(2008)\citenamefont
  {Akhmerov}, \citenamefont {Bardarson}, \citenamefont {Rycerz},\ and\
  \citenamefont {Beenakker}}]{Beenakker08}%
  \BibitemOpen
  \bibfield  {author} {\bibinfo {author} {\bibfnamefont {A.~R.}\ \bibnamefont
  {Akhmerov}}, \bibinfo {author} {\bibfnamefont {J.~H.}\ \bibnamefont
  {Bardarson}}, \bibinfo {author} {\bibfnamefont {A.}~\bibnamefont {Rycerz}},\
  and\ \bibinfo {author} {\bibfnamefont {C.~W.~J.}\ \bibnamefont {Beenakker}},\
  }\bibfield  {title} {\bibinfo {title} {Theory of the valley-valve effect in
  graphene nanoribbons},\ }\href {https://doi.org/10.1103/PhysRevB.77.205416}
  {\bibfield  {journal} {\bibinfo  {journal} {Phys. Rev. B}\ }\textbf {\bibinfo
  {volume} {77}},\ \bibinfo {pages} {205416} (\bibinfo {year}
  {2008})}\BibitemShut {NoStop}%
\bibitem [{\citenamefont {Silaev}(2021)}]{Silaev}%
  \BibitemOpen
  \bibfield  {author} {\bibinfo {author} {\bibfnamefont {M.}~\bibnamefont
  {Silaev}},\ }\href@noop {} {\bibinfo {title} {Multiple optical gaps and laser
  with magnonic pumping in 2{D} {I}sing superconductors}} (\bibinfo {year}
  {2021}),\ \Eprint {https://arxiv.org/abs/arXiv:2111.03623} {arXiv:2111.03623}
  \BibitemShut {NoStop}%
\bibitem [{\citenamefont {Pandey}\ \emph {et~al.}(2021)\citenamefont {Pandey},
  \citenamefont {Danneau},\ and\ \citenamefont {Beckmann}}]{Pandey21}%
  \BibitemOpen
  \bibfield  {author} {\bibinfo {author} {\bibfnamefont {P.}~\bibnamefont
  {Pandey}}, \bibinfo {author} {\bibfnamefont {R.}~\bibnamefont {Danneau}},\
  and\ \bibinfo {author} {\bibfnamefont {D.}~\bibnamefont {Beckmann}},\
  }\bibfield  {title} {\bibinfo {title} {Ballistic graphene {C}ooper pair
  splitter},\ }\href {https://doi.org/10.1103/PhysRevLett.126.147701}
  {\bibfield  {journal} {\bibinfo  {journal} {Phys. Rev. Lett.}\ }\textbf
  {\bibinfo {volume} {126}},\ \bibinfo {pages} {147701} (\bibinfo {year}
  {2021})}\BibitemShut {NoStop}%
\bibitem [{\citenamefont {Cioni}\ and\ \citenamefont
  {Taddei}(2025)}]{Taddei25}%
  \BibitemOpen
  \bibfield  {author} {\bibinfo {author} {\bibfnamefont {F.}~\bibnamefont
  {Cioni}}\ and\ \bibinfo {author} {\bibfnamefont {F.}~\bibnamefont {Taddei}},\
  }\href@noop {} {\bibinfo {title} {High-performance {A}ndreev
  interferometer-based electronic coolers}} (\bibinfo {year} {2025}),\ \Eprint
  {https://arxiv.org/abs/arXiv:2503.17054} {arXiv:2503.17054} \BibitemShut
  {NoStop}%
\bibitem [{\citenamefont {Kopnin}(2001)}]{noneqSC}%
  \BibitemOpen
  \bibfield  {author} {\bibinfo {author} {\bibfnamefont {N.}~\bibnamefont
  {Kopnin}},\ }\href@noop {} {\emph {\bibinfo {title} {Theory of Nonequilibrium
  Superconductivity}}}\ (\bibinfo  {publisher} {Oxford University Press},\
  \bibinfo {year} {2001})\BibitemShut {NoStop}%
\end{thebibliography}%

\end{document}